\documentclass{IEEEtran4PSCC}
\usepackage{amsmath}
\usepackage{amsfonts}
\usepackage{amssymb}
\usepackage{bm}
\usepackage[colorlinks=false,hidelinks]{hyperref}  
\usepackage[capitalize]{cleveref}
\crefname{equation}{}{}
\usepackage{graphicx}
\usepackage{doi}

\usepackage{glossaries}
\usepackage{xcolor}

\usepackage{algorithm}
\usepackage[noend]{algpseudocode}
\usepackage{nicefrac}
\usepackage{booktabs}  
\usepackage{siunitx}
\usepackage{dirtytalk}
\usepackage{multicol}
\usepackage[shortlabels]{enumitem}
\usepackage{multirow}
\usepackage{tabularx}

\usepackage{tikz}
\usepackage{pgfplots}
\pgfplotsset{compat=1.18}
\usepgfplotslibrary{groupplots}

\DeclareSIUnit{\pu}{p.u.}

\DeclareMathOperator{\PINN}{PINN}

\newcommand{\ddt}{\ensuremath{\frac{d}{dt}}}
\newcommand{\norm}[2]{\ensuremath{\left\lVert#1\right\rVert_{#2}}}

\newcommand{\fupdate}{\ensuremath{\bm{f}}}
\newcommand{\fupdateof}[1]{\ensuremath{\fupdate\left(#1\right)}}
\newcommand{\fupdateiof}[2]{\ensuremath{\fupdate_{#2}\left(#1\right)}}

\newcommand{\gupdate}{\ensuremath{\bm{g}}}
\newcommand{\gupdateof}[1]{\ensuremath{\gupdate\left(#1\right)}}

\newcommand{\xstate}{\ensuremath{\bm{x}}}
\newcommand{\ystate}{\ensuremath{\bm{y}}}

\newcommand{\xinitial}{\ensuremath{\bm{x}_0}}
\newcommand{\xinitiali}[1]{\ensuremath{\bm{x}_{0,#1}}}
\newcommand{\xstatei}[1]{\ensuremath{\bm{x}_{#1}}}
\newcommand{\ustatei}[1]{\ensuremath{\bm{u}_{#1}}}
\newcommand{\Rdim}[1]{\ensuremath{\mathbb{R}^{#1}}}
\newcommand{\Cdim}[1]{\ensuremath{\mathbb{C}^{#1}}}

\newcommand{\xstatehat}{\ensuremath{\hat{\bm{x}}}}
\newcommand{\xstatehati}[1]{\ensuremath{\hat{\bm{x}}_{#1}}}
\newcommand{\ystatehat}{\ensuremath{\hat{\bm{y}}}}

\newcommand{\Ycomplex}{\ensuremath{\Bar{\bm{Y}}}}
\newcommand{\realof}[1]{\ensuremath{\Re\left(#1\right)}}
\newcommand{\imagof}[1]{\ensuremath{\Im\left(#1\right)}}

\newcommand{\zeroDim}[1]{\ensuremath{\bm{0}_{#1}}}

\newcommand{\Vcomplex}{\ensuremath{\Bar{\bm{v}}}}
\newcommand{\Vcomplexhat}{\ensuremath{\hat{\Bar{\bm{v}}}}}
\newcommand{\IcomplexComponent}{\ensuremath{\Bar{\bm{i}}^C}}
\newcommand{\IcomplexNetwork}{\ensuremath{\Bar{\bm{i}}^N}}
\newcommand{\IcomplexComponenti}[1]{\ensuremath{\Bar{i}_{#1}^C}}
\newcommand{\Vcomplexi}[1]{\ensuremath{\Bar{v}_{#1}}}
\newcommand{\Vcomplexihat}[1]{\ensuremath{\hat{\Bar{v}}_{#1}}}

\newcommand{\IcomplexComponentHat}{\ensuremath{\hat{\Bar{\bm{i}}}^C}}
\newcommand{\IcomplexComponentHati}[1]{\ensuremath{\hat{\Bar{i}}_{#1}^C}}
\newcommand{\IcomplexNetworkHat}{\ensuremath{\hat{\Bar{\bm{i}}}^N}}
\newcommand{\IcomplexNetworkHati}[1]{\ensuremath{\hat{\Bar{i}}_{#1}^N}}
\newcommand{\Vparams}{\ensuremath{\bm{\Xi}}}
\newcommand{\Vparamsi}[1]{\ensuremath{\bm{\Xi}_{#1}}}
\newcommand{\Vparamsiter}[1]{\ensuremath{\bm{\Xi}^{(#1)}}}
\newcommand{\Vparamsiteri}[2]{\ensuremath{\bm{\Xi}_{#2}^{(#1)}}}

\newcommand{\residualError}{\ensuremath{\bm{\rho}}}
\newcommand{\Jacobian}{\ensuremath{\bm{J}}}

\newcommand{\normtwosquared}[1]{\ensuremath{\left\lVert#1\right\rVert_{2}^{2}}}
\newcommand{\sizeof}[1]{\ensuremath{\left\lvert#1\right\rvert}}
\newcommand{\oneoversizeof}[1]{\ensuremath{\frac{1}{\left\lvert#1\right\rvert}}}

\newcommand\revision[2]{\iftoggle{showRevision}{\textcolor{black}{#2}}{#1}}

\newacronym{PINN}{PINN}{Physics-Informed Neural Network}
\newacronym{RK}{RK}{Runge-Kutta}
\newacronym{DAE}{DAE}{Differential-Algebraic Equation}
\newacronym{DE}{DE}{Differential Equation}

\newacronym{NN}{NN}{Neural Network}
\newacronym{AD}{AD}{Automatic Differentiation}
\newacronym{TDS}{TDS}{Time-Domain Simulation}

\makeatletter
\let\old@ps@headings\ps@headings
\let\old@ps@IEEEtitlepagestyle\ps@IEEEtitlepagestyle
\def\psccfooter#1{%
    \def\ps@headings{%
        \old@ps@headings%
        \def\@oddfoot{\strut\hfill#1\hfill\strut}%
        \def\@evenfoot{\strut\hfill#1\hfill\strut}%
    }%
    \def\ps@IEEEtitlepagestyle{%
        \old@ps@IEEEtitlepagestyle%
        \def\@oddfoot{\strut\hfill#1\hfill\strut}%
        \def\@evenfoot{\strut\hfill#1\hfill\strut}%
    }%
    \ps@headings%
}
\makeatother

\psccfooter{%
        \parbox{\textwidth}{\hrulefill \\ \small{23rd Power Systems Computation Conference} \hfill \begin{minipage}{0.2\textwidth}\centering \vspace*{4pt} \includegraphics[scale=0.06]{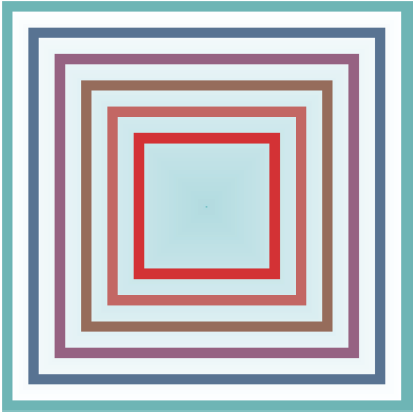}\\\small{PSCC 2024} \end{minipage} \hfill \small{Paris, France --- June 4 -- 7, 2024}}%
}

\preto\subequations{\ifhmode\unskip\fi}

\newtoggle{showRevision}
\toggletrue{showRevision}

\begin{document}
\bstctlcite{IEEEexample:BSTcontrol}
%
\title{PINNSim: A Simulator for Power System Dynamics\\based on Physics-Informed Neural Networks}

\author{
\IEEEauthorblockN{Jochen Stiasny, Spyros Chatzivasileiadis}
\IEEEauthorblockA{Division for Power and Energy Systems \\
Technical University of Denmark\\
Kgs. Lyngby, Denmark\\
\{jbest, spchatz\}@dtu.dk}
\and
\IEEEauthorblockN{Baosen Zhang}
\IEEEauthorblockA{Electrical and Computer Engineering\\
University of Washington\\
Seattle, WA 98195\\ zhangbao@uw.edu}
}

\maketitle

\begin{abstract}
The dynamic behaviour of a power system can be described by a system of differential-algebraic equations. Time-domain simulations are used to simulate the evolution of these dynamics. They often require the use of small time step sizes and therefore become computationally expensive. To accelerate these simulations, we propose a simulator -- PINNSim -- that allows to take significantly larger time steps. It is based on Physics-Informed Neural Networks (PINNs) for the solution of the dynamics of single components in the power system. To resolve their interaction we employ a scalable root-finding algorithm. We demonstrate PINNSim on a 9-bus system and show the increased time step size compared to a trapezoidal integration rule. We discuss key characteristics of PINNSim and important steps for developing PINNSim into a fully fledged simulator. As such, it could offer the opportunity for significantly increasing time step sizes and thereby accelerating time-domain simulations.
\end{abstract}


\thanksto{\noindent This work was supported by the ERC Project VeriPhIED, funded by the European Research Council, Grant Agreement No: 949899.}

\section{Introduction}\label{sec:introduction}

Differential equations form a centre piece in the modelling of the dynamic behaviour of power systems. They provide very widely applicable component and system models, however, their solution requires numerical integration methods. These tools for \glspl{TDS} constitute a workhorse of power system analysis and can reliably deliver accurate solutions. The used algorithms are scalable, versatile, and applicable without case-specific modifications, but they incur a high computational burden \cite{stott_power_1979}. Efforts around improved solvers and parallelisation can reduce this burden, see \cite{liu_solving_2020, gurrala_parareal_2016, aristidou_time-domain_2015} for an overview. However, a fundamental problem is the small time step size that is required by the schemes to ensure accurate results and numerical stability. Hence many time steps and function evaluations are necessary.

The proposition of \gls{NN}-based approaches \cite{lagaris_artificial_1998}, and in particular \glspl{PINN} \cite{raissi_physics-informed_2018}, has very different characteristics. In the training process, \glspl{PINN} learn the solution to the provided differential equations and subsequently provide fast and sufficiently accurate solutions over long time intervals -- addressing two main issues of conventional numerical integration methods.
\revision{
\glspl{PINN} have been adopted to power system dynamics in \cite{misyris_physics-informed_2020} to predict single machine dynamics. In \cite{moya_approximating_2023}, the related methodology of operator learning is also applied to single machine dynamics. The applicability of learning-based methods to multi-machine systems is explored in \cite{stiasny_physics-informed_2023} and with alternative architectures in \cite{cui_predicting_2021,moya_dae-pinn_2023}. The speed advantage persists and the accuracy remains sufficiently high for the trained setup. However, the common challenge lies in achieving generalisation, i.e., being able to apply the same \gls{NN} for an unseen task without retraining. Such flexibility would make \gls{NN}-based \gls{TDS} approaches comparable with conventional numerical integration methods. However, in the case of multi-machine systems, this requires the generalisation to a very large variety of setups. As a consequence, the dimensionality of the associated learning tasks increases dramatically with the system size and thereby becomes very or even too costly to learn \cite{stiasny_physics-informed_2023-1}.
}{
The authors in \cite{legaard_constructing_2023} refer to these \glspl{NN} for simulating dynamical systems as a direct-solution model.}

\revision{}{
As such, \glspl{PINN} have been adopted to power system dynamics in \cite{misyris_physics-informed_2020} to predict single machine dynamics. In a similar setup, the related methodology of operator learning is applied \cite{moya_approximating_2023}. Other learning-based methods for simulating dynamics -- see \cite{legaard_constructing_2023} for an overview -- have also been explored in the context of single components, e.g. in \cite{xiao_feasibility_2022}. It is an open question if the promising results for single components can also be achieved when predicting the dynamics of multi-machine systems. For PINNs, \cite{stiasny_physics-informed_2023,moya_dae-pinn_2023} explored this question for a 9, 11, and 39-bus system. The authors in \cite{cui_predicting_2021} predict power system transients with an alternative architecture which relies on a transformation into the frequency domain. The effect of generator outages is predicted in \cite{roberts_continuous-time_2022} and \cite{li_machine-learning-based_2020} estimates system dynamics after line faults.} 

\revision{}{In the above cases, the speed advantage of learning-based methods persists and the accuracy remains sufficiently high for the trained setup. However, all approaches require assumptions about the operating conditions, such as the network topology and the set-points, as well as the disturbance type and location during the dataset generation and the training of the model. As a result these learning-based methods cannot be reliably applied to unseen conditions or faults. If one attempted to include all possible conditions and faults in the learning stage, the resulting problem will not be tractable as the required dataset becomes too high-dimensional \cite{stiasny_physics-informed_2023-1}. Hence, attempts to scale \gls{NN}-based \gls{TDS} approaches need to focus on limiting the assumptions that are required for the learning task such as operating conditions and disturbance characteristics.
}

\revision{To nonetheless harness the advantages of \glspl{PINN}, i.e., fast solutions for large time steps, we propose \textit{PINNSim}. This algorithm allows a modular integration of \glspl{PINN} into \glspl{TDS} for multi-machine systems. We learn the solution of all dynamic components in the system separately, i.e., with entirely independent \glspl{PINN}, and before executing any simulation. Thereby, the learning tasks are of relatively low dimensionality and hence remain tractable. To perform a \gls{TDS}, we solely need to determine the interaction between the components. To this end, we utilise a root-finding algorithm similar to conventional \gls{TDS} algorithms; at this stage, no learning is required. As a result, the scalability and flexibility of PINNSim become decoupled from the difficulties of learning high-dimensional problems. At the same time, we still can benefit from fast and accurate solutions for larger time step sizes due to the use of \glspl{PINN}.
}{
To that end, we propose the simulator \textit{PINNSim}. We first learn the solutions of all dynamic components in the system separately, i.e., with entirely independent \glspl{PINN}. Thereby, the learning tasks are of relatively low dimensionality and hence remain tractable. Meanwhile, the use of \glspl{PINN} allows us to benefit from fast solutions for large time steps. To then perform a \gls{TDS} of the multi-machine system, we need to determine the interactions between the components. We achieve this by applying a root-finding algorithm similar to conventional \gls{TDS} algorithms; at this stage, no learning is required. As a result, the scalability and flexibility of PINNSim become decoupled from the difficulties of learning high-dimensional problems. At the same time, we still can benefit from fast and accurate solutions for larger time step sizes due to the use of \glspl{PINN}.}

In \cref{sec:concept}, we present the conceptual motivation behind PINNSim. In \cref{sec:methodolgy}, we introduce the power system formulation and the methodology of PINNSim. \Cref{sec:case_study} describes the \revision{case study}{setup of the numerical experiments to demonstrate a proof of concept} and in \cref{sec:results}, we discuss the results by highlighting key characteristics of PINNSim. \Cref{sec:discussion} discusses the steps for developing PINNSim from a proof of concept to a fully fledged simulator. \Cref{sec:conclusion} concludes.
\section{Concept}\label{sec:concept}

To facilitate the presentation of the methodology in \cref{sec:methodolgy}, we want to first describe the conceptual idea behind PINNSim. It originates from the problem that we face when solving \glspl{DAE}, a form of differential equations where a set of algebraic equations constraints the differential equations. It formulates as
\begin{subequations}\label{eq:DAE_general}%
\begin{align}
    \ddt \xstate &= \fupdateof{\xstate, \ystate}\\
    \bm{0} &= \gupdateof{\xstate, \ystate}
\end{align}
\end{subequations}
and we refer to $\xstate(t)$ as the differential variables and to $\ystate(t)$ as the algebraic variables. The update function \fupdateof{\xstate, \ystate} and the algebraic relationship  \gupdateof{\xstate, \ystate} govern the dynamics of the system\footnote{For notational clarity, we formulate an autonomous, unforced system. The conceptual idea can also accommodate non-autonomous and forced systems.}. Our interest focuses on the particular form of index-1 \glspl{DAE} or semi-analytical \glspl{DAE} \cite{brenan_numerical_1995}. This form implies that \gupdate{} can be differentiated once with respect to time $t$ which is possible when $\frac{\partial \gupdate{}}{\partial \ystate}$ is non-singular. Based on \cref{eq:DAE_general}, we could describe the temporal evolution of \xstate{} and \ystate{} as 
\begin{subequations}\label{eq:DAE_general_integral}
\begin{align}
    \xstate(t) &= \xinitial + \int_{t_0}^t \fupdateof{\xstate, \ystate} d\tau\\
    \ystate(t) &= \gupdate'(\xstate)
\end{align}
\end{subequations}
where $\xinitial = \xstate(t_0)$ represents the initial condition and $\gupdate'(\xstate)$ describes the solution to the algebraic equations given \xstate{}. However, usually no analytical expression describes the evolution $\xstate(t)$ and $\ystate(t)$ for a given \xinitial{}. Hence, we revert to numerical integration methods to obtain an approximate solution. 

To resolve the non-trivial integration operation and the implicit relationship \gupdate{}, numerical schemes often restrict the functional form of \xstate{} to a certain approximation \xstatehat{}. For instance, \gls{RK} methods assume a polynomial form of $\xstatehat{}(t) = \xinitial + a_1 (t-t_0) +  a_2 (t-t_0)^2 + \dots$ as they match the Taylor expansion up to a certain degree by construction. The different \gls{RK} schemes prescribe the order of the scheme and the computation of the coefficients. When algebraic variables are present, they have to be interfaced with the approximation of the differential variable to incorporate their interaction with each other. Simultaneous and partitioned integration methods describe such routines \cite{stott_power_1979} and are used in many variations. The accuracy of these constructions is dependent on the order of the used integration scheme and potential \say{interface} errors. These considerations and aspects of numerical stability limit the usable time step size. 

By choosing a different functional form for \xstatehat{}, like Fourier-series based or around the Adomian decomposition \cite{adomian_review_1988}, the practical time step size might be increased. Due to their high flexibility of their functional form, \glspl{PINN}, and \glspl{NN} in general as suggested in \cite{lagaris_artificial_1998}, can allow significantly larger time steps. In fact, we can even choose a functional form of $\ystatehat{}(t)$ and approximate $\xstatehat{} = \PINN(t, \ystatehat{})$ in dependence of it. However, this great approximation flexibility of \glspl{PINN} comes with the challenge of generalising well across the entire domain of interest, i.e., being accurate over the entire domain and not only on the training dataset. When the dimensionality of \xstate{} and \ystate{} in \cref{eq:DAE_general} increases, the training of a \gls{PINN} to approximate a wide range of solutions becomes increasingly difficult and eventually intractable. With PINNSim, we avoid this problem by exploiting the structure of the power system specific \glspl{DAE}. The structure allows a decomposition of \cref{eq:DAE_general} into multiple smaller sub-problems which remain tractable from a learning perspective. At the same time, the use of \glspl{PINN} allows for large and accurate time steps of these sub-problems. To simulate the entire system, we need to align the sub-problems' solutions by enforcing the algebraic relationship \gupdateof{\xstate{}, \ystate{}}.

\section{Methodology}\label{sec:methodolgy}

This section describes the problem formulation and its decomposition for PINNSim in \cref{subsec:power_system_DAEs}. We then describe the main elements of the algorithm in \cref{subsec:voltage_profile,subsec:PINN_approximation,subsec:voltage_profile_update} and lastly the entire algorithm in \cref{subsec:full_algorithm}. 

\subsection{Problem formulation and solution approach}\label{subsec:power_system_DAEs}
The key relation for power system dynamics is described by the current balance, i.e., Kirchhoff's current law, and it needs to hold at all times $t$ and for each of the $n$ buses in the network. We distinguish between the complex current injections stemming from the network $\IcomplexNetwork{} \in \Cdim{n}$ and from connected components $\IcomplexComponent{} \in \Cdim{n}$. The former can be described by the algebraic relationship (neglecting electro-magnetic transients) 
\begin{align}
    \IcomplexNetwork{} = \Ycomplex \Vcomplex.
\end{align}
where $\Vcomplex \in \Cdim{n}$ represents the complex voltages at the buses and $\Ycomplex \in \Cdim{n\times n}$ the complex admittance matrix. The component current injections $\IcomplexComponenti{i}$ at bus $i$ can stem from a \textit{dynamic component}, i.e., their behaviour is governed by differential equations,
\begin{subequations}
\begin{align}
    \IcomplexComponenti{i} &= h_i(\xstatei{i}, \Vcomplexi{i})\label{eq:h_update_power_system}\\
    \ddt \xstatei{i} &= \fupdateiof{\xstatei{i}, \Vcomplexi{i},\ustatei{i}}{i}\label{eq:f_update_power_system}
\end{align} 
\end{subequations}
where the current injection depends on the state vector $\xstatei{i} \in \Rdim{p}$ and the voltage \Vcomplexi{i} of the bus $i$ to which the component is connected\footnote{For notational ease, we assume that component $i$ is connected to bus $i$. If component $i$ was connected to bus $j$, \Vcomplexi{i} would be replaced by \Vcomplexi{j}. If a component is connected to multiple buses, all of the corresponding voltages will be included in $h_i(\cdot)$ and \fupdateiof{\cdot}{i} in \cref{eq:f_update_power_system,eq:h_update_power_system}.}. The update function $\fupdate{}_i$ can furthermore depend on control inputs \ustatei{i}. For \textit{static components} the current injections becomes a function of the local voltage $\IcomplexComponenti{i} = h_i(\Vcomplexi{i})$. If no component is connected to bus $i$, then $\IcomplexComponenti{i} = 0$. \revision{If multiple components were connected to the same bus, the current injections would be summed up.}{If multiple components, indexed by $k$, were connected to the same bus, their current injections $\Bar{i}_{i}^{C} = \sum_k \Bar{i}_{i, k}^{C}$ are summed up and the resulting current injection then depends on the states of the connected dynamic components $\xstatei{i,k}$ and the local voltage \Vcomplexi{i}.} 
\begin{figure}[!t]
    \centering
    \includegraphics[width=\linewidth]{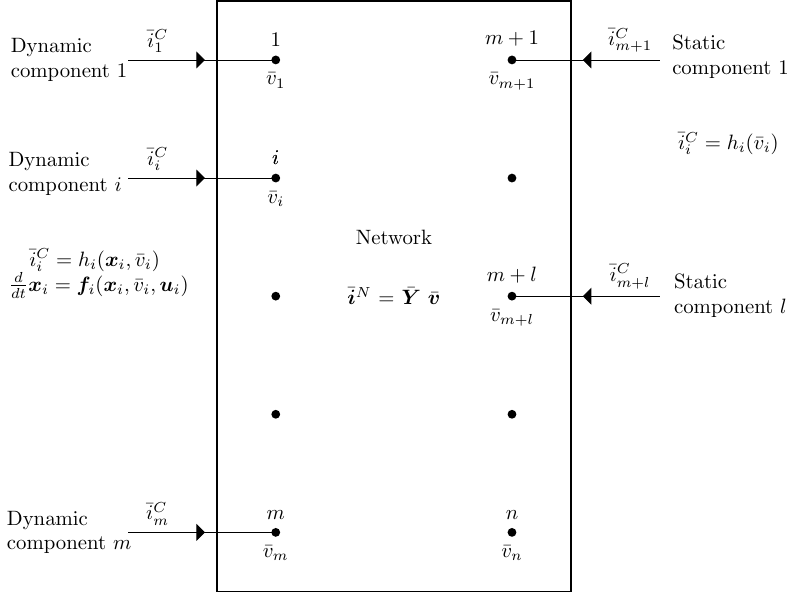}
    \caption{Structure of the \glspl{DAE} that govern the power system dynamics with current injections \IcomplexComponent{} from the components, e.g., generators and loads, and current flows in the network \IcomplexNetwork{}. Adapted from \cite{sauer_power_1998}.}
    \label{fig:power_system_structure}
\end{figure}

\Cref{fig:power_system_structure} schematically depicts this structure with $m$ dynamic components, $l$ static component, and $n$ buses. When we formulate the current balance
\begin{align}\label{eq:current_balance}
    \zeroDim{} = \IcomplexComponent{}(\xstate(t), \Vcomplex(t)) - \IcomplexNetwork{}(\Vcomplex(t)),
\end{align}
and consider the system's structure, we note that the component currents \IcomplexComponent{} have only a \textit{local} dependency, i.e., to their corresponding state vectors \xstatei{i} and the local voltage \Vcomplexi{i}. Their interaction arises from the coupling through the network currents \IcomplexNetwork{}, which in turn only depends on all complex voltages $\Vcomplex$.

PINNSim exploits this structure as follows: We assume a voltage evolution $\Vcomplexhat{}(t)$ -- we denote all approximations with the hat symbol $\hat{}$ -- based on which we evaluate the resulting network currents \IcomplexNetworkHat{} and the component currents \IcomplexComponentHat{}. For the latter, we furthermore require the state evolution \xstatehati{i} of each component; we approximate the necessary temporal integration with \glspl{PINN}. When comparing the resulting currents
\begin{align}
    \IcomplexComponentHat{}\left(\xstatehat(t), \Vcomplexhat(t)\right) - \IcomplexNetworkHat{}\left(\Vcomplexhat(t)\right)
\end{align}
the result will not equal $\bm{0}$ by default, i.e., the current balance \cref{eq:current_balance} is not obeyed. However, the mismatch indicates the quality of the approximation. By adjusting the assumed voltage evolution $\Vcomplexhat{}(t)$, we iteratively reduce this mismatch. The resulting voltage $\Vcomplexhat{}(t)$ and state $\xstatehat{}(t)$ evolutions constitute the solution of PINNSim to the system of \glspl{DAE}.

\subsection{Parametrisation of the voltage evolution $\Vcomplexihat{i}(t)$}\label{subsec:voltage_profile}

We describe the evolution of the complex voltage in polar form $\Vcomplexi{i}(t) = V_{i}(t) e^{j\theta_i(t)}$. To approximate $\Vcomplexi{i}(t)$, we express the voltage magnitude $V_i$ and the voltage angle $\theta_i$ as a power-series with respect to time up to power $r$ as suggested in \cite{wang_timepower_2019} 
\begin{align}
     \Vcomplexihat{i}(t) &= \left(\sum_{k=0}^{r} V_{k, i} (t-t_0)^k \right) e^{j\left(\sum_{k=0}^{r} \theta_{k, i} (t-t_0)^k \right)}.\label{eq:voltage_parametrisation}
\end{align}
The coefficients $V_{0, i}, V_{1, i}, \dots, V_{r, i}$ and $\theta_{0, i}, \theta_{1, i}, \dots, \theta_{r, i}$ form the parameters which we will later on update to improve the approximation. The vector $\Vparams{}_i$ collects all parameters at bus $i$
\begin{align}
    \Vparams{}_i = \begin{bmatrix}
      V_{0, i} & \theta_{0, i} & \hdots & V_{r, i}, \theta_{r, i}
    \end{bmatrix}, \quad \Vparams{}_i \in \Rdim{2(r+1)}.
\end{align}
This parametrisation is repeated for all $n$ buses in the system. We concatenate all $\Vparams{}_i$ in the vector $\Vparams{} \in \Rdim{2 (r+1) n}$.

\subsection{Solving component dynamics with PINNs}\label{subsec:PINN_approximation}
The exact solution for the evolution of the differential variables $\bm{x}_i(t)$ of a single dynamic component $i$ can be obtained by integration of \cref{eq:f_update_power_system}
\begin{align}
    \xstatei{i}(t) &= \bm{x}_{0, i} + \int_{t_0}^{t} \bm{f}_i\left(\xstatei{i}(\tau), \Vcomplexi{i}(\tau), \ustatei{i}\right) d\tau. \label{eq:x_integral}
\end{align}
As there usually exists no explicit analytical solution to \labelcref{eq:x_integral}, we need to approximate the solution. As motivated earlier, using \glspl{PINN} allows us to accurately approximate \labelcref{eq:x_integral} over large time steps $\Delta t=t-t_0$ while being fast to evaluate. The approximation \revision{will be dependent}{depends} on the time step size $\Delta t$, the initial state $ \bm{x}_{0, i}$, potentially the control input \ustatei{i}, and the voltage profile \Vcomplexi{i}. As we cannot train the \gls{PINN} with an arbitrary \Vcomplexi{i}, we restrict the learning to the same form as in \cref{eq:voltage_parametrisation}, i.e., $\Vcomplexi{i} \approx \Vcomplexihat{i}(\Delta t, \Vparamsi{i})$. Hence, the approximation of \cref{eq:x_integral} will \revision{depend also}{also depend} on $\Vparamsi{i}$
\begin{align}
    \xstatehati{i}(t) = \PINN(\Delta t, \bm{x}_{0, i}, \Vparamsi{i}, \bm{u}_i).
\end{align}
The functional form of the \gls{PINN} is a standard feed-forward \gls{NN} with $K$ hidden layers \revision{parameterised}{parametrised} by the weight matrices $\bm{W}^{k}$ and bias vectors $\bm{b}^{k}$ and a non-linear activation function $\sigma$
\begin{subequations}
\begin{align}
    \bm{z}_0 &= [\Delta t, \bm{x}_{0,i}, \Vparamsi{i}, \bm{u}_i] \label{eq:NN_input}\\
    \bm{z}_{k+1} &= \sigma\left( \bm{W}^{k+1} \bm{z}_{k} + \bm{b}^{k+1}\right), \,\forall k = 0, 1, \ldots, K-1\label{eq:NN_hidden_layers}\\
    \xstatehati{i} &= \xinitiali{i} + \Delta t (\bm{W}^{K} \bm{z}_K + \bm{b}^{K}). \label{eq:NN_output}
\end{align}
\end{subequations}
We adjust the last layer to enforce that the initial condition $\bm{x}_{0,i}$ is met if $\Delta t=0$, i.e., $t = t_0$. Thereby we ensure numerical consistency and can improve inaccuracies related to recursive application of the \gls{PINN} \cite{lagaris_artificial_1998}. 
To train a \gls{PINN}, we optimise
\begin{align}
\min_{\bm{W}^{1}, \bm{b}^{1}, \dots, \bm{W}^{K}, \bm{b}^{K}} \quad \mathcal{L}_{x} + \alpha \mathcal{L}_{c},
\end{align}
where $\mathcal{L}_x$ evaluates the prediction error based on a dataset $\mathcal{D}$ of simulated data points 
\begin{align}
\mathcal{L}_{x} = \oneoversizeof{\mathcal{D}}\sum_{j=1}^{\sizeof{\mathcal{D}}} \normtwosquared{\xstatei{i}^{(j)} - \xstatehati{i}^{(j)}}
\end{align}
and a physics-based loss term $\mathcal{L}_{c}$ evaluated on \textit{collocation} points which do not require any simulation
\begin{align}    
\mathcal{L}_{c} = &\oneoversizeof{\revision{\mathcal{D}}{\mathcal{D}_c}}\sum_{j=1}^{\sizeof{\mathcal{D}_c}} \normtwosquared{\ddt \hat{\bm{x}}_{i}^{(j)} - \fupdateof{\xstatehati{i}^{(j)}, \revision{\Vcomplexi{i}^{(j)}}{\Vcomplexihat{i}^{(j)}}, \ustatei{i}^{(j)}}}.
\end{align}
The two loss terms are weighted with the hyperparameter $\alpha$. For a more detailed explanation of \glspl{PINN}, we refer to \cite{stiasny_physics-informed_2023}. Based on the evolution of the state $\xstatehati{i}(t)$ and the voltage $\Vcomplexihat{i}(t)$, we can calculate the current injection \IcomplexComponentHati{i} by evaluating
\begin{align}
    \IcomplexComponentHati{i}(t) = h_i(\xstatehati{i}(t), \Vcomplexihat{i}(t)).
\end{align}
For each of the $m$ dynamic components, we will train a separate \gls{PINN} and all training \revision{can be can be}{can be} performed in advance of executing PINNSim.

\subsection{Updating the voltage profile}\label{subsec:voltage_profile_update}

The approximations $\Vcomplexihat{i}(t)$ and $\xstatehati{i}(t)$ are both continuous functions and therefore the same holds for $\IcomplexComponentHati{i}(t)$ and $\IcomplexNetworkHati{i}(t)$. The current balance \cref{eq:current_balance} will only be fulfilled, if $\IcomplexComponentHati{i}(t) = \IcomplexNetworkHati{i}(t)$ for the entire time step and at all buses. We express this notion by requiring that the norm $\norm{\IcomplexComponentHati{i}-\IcomplexNetworkHati{i}}{}$ shall be 0. This norm
\begin{align}\label{eq:metric_definition}
    \norm{\IcomplexComponentHati{i}-\IcomplexNetworkHati{i}}{} = \sqrt{ \langle \IcomplexComponentHati{i}-\IcomplexNetworkHati{i}, \IcomplexComponentHati{i}-\IcomplexNetworkHati{i}\rangle}
\end{align}
is induced by the inner product $\langle \Bar{a}, \Bar{b}\rangle$ between two complex functions $\Bar{a}(t), \Bar{b}(t)$ over the interval $[t_0, t_0 + \Delta t]$
\begin{align}
    \langle \Bar{a}, \Bar{b}\rangle &= \int_{t_0}^{t_0 + \Delta t} \Bar{a}(t) \, \Bar{b}^*(t) \, dt.
\end{align}
As \IcomplexComponentHati{i} and \IcomplexNetworkHati{i} depend on the parametrised voltages $\Vcomplexhat{}(t, \Vparams)$, we subsequently aim to find a parametrisation \Vparams{} that minimises $\norm{\IcomplexComponentHati{i}-\IcomplexNetworkHati{i}}{}$. To this end, we formulate the following optimisation problem for the entire system as the sum of the norms (we square  \cref{eq:metric_definition} to avoid the calculation of the square root)
\begin{align}\label{eq:objective_voltage_update_exact}
    \min_{\Vparams} \sum_{i=1}^n \norm{\IcomplexComponentHati{i} - \IcomplexNetworkHati{i}}{}^2.
\end{align}
The above expression involves the integration
\begin{align*}
    \norm{\IcomplexComponentHati{i} - \IcomplexNetworkHati{i}}{}^2 &=\int_{t_0}^{t_0 + \Delta t} \realof{\IcomplexComponentHati{i} - \IcomplexNetworkHati{i}}^2 + \imagof{\IcomplexComponentHati{i} - \IcomplexNetworkHati{i}}^2 dt
\end{align*}
which we approximate with the Midpoint rule, i.e., we split the interval $[t_0, t_0 + \Delta t]$ into $s$ equally sized intervals of width $\frac{\Delta t}{s}$ and sum the function value at the middle of these intervals 
\begin{align}\label{eq:objective_approximation}
\begin{split}
    \norm{\IcomplexComponentHati{i} - \IcomplexNetworkHati{i}}{}^2 \approx \frac{\Delta t}{s}\sum_{j=1}^s &\; \realof{\IcomplexComponentHati{i}(t_j) - \IcomplexNetworkHati{i}(t_j)}^2 \\&+ \imagof{\IcomplexComponentHati{i}(t_j) - \IcomplexNetworkHati{i}(t_j)}^2.
\end{split}
\end{align}
Thereby, we can approximate the optimisation in \cref{eq:objective_voltage_update_exact} as
\begin{align}
    \min_{\Vparams} \frac{\Delta t}{s} \residualError{}^\top \residualError{}
\end{align}
where \residualError{} collects all summands in a vector
\begin{align}
\begin{split}
    \residualError{} &= \begin{bmatrix} \realof{\IcomplexComponentHati{1}(t_1) - \IcomplexNetworkHati{1}(t_1)} \\ \imagof{\IcomplexComponentHati{1}(t_1) - \IcomplexNetworkHati{1}(t_1)} \\ \vdots \\\realof{\IcomplexComponentHati{n}(t_s) - \IcomplexNetworkHati{n}(t_s)} \\ \imagof{\IcomplexComponentHati{n}(t_s) - \IcomplexNetworkHati{n}(t_s)} \end{bmatrix}, \quad \residualError{} \in \Rdim{2ns}
\end{split}
\end{align}

We solve the above non-linear least square problem by iteratively updating the parameters $\Vparamsiter{k+1} = \Vparamsiter{k} + \Delta \Vparams$. To determine $\Delta \Vparams$, we compute the Jacobian  
\begin{align}\label{eq:jacobian_least_square}
    \bm{J} &= \frac{\partial \residualError{}}{\partial \Vparams}\Bigr|_{\Vparamsiter{k}}, \quad \bm{J} \in \Rdim{2ns\times2(r+1)n}
\end{align}
at the values \Vparamsiter{k} and solve the linear problem
\begin{align}\label{eq:residual_least_square}
    \left(\Jacobian^\top \Jacobian\right) \Delta \Vparams = - \Jacobian^\top \residualError{}.
\end{align}
To calculate \Jacobian{} in \cref{eq:jacobian_least_square}, we use \gls{AD} \cite{baydin_automatic_2018} as all computations for \IcomplexComponentHat{} and \IcomplexNetworkHat{} are expressed as explicit functions. While the size of \Jacobian{} can become large, it has a very sparse structure that can be exploited in its construction and when solving \cref{eq:residual_least_square} -- herein lies the key to the scalability of PINNSim as it closely resembles the structure of conventional integration schemes such as the trapezoidal method.

\subsection{PINNSim: The full time-stepping simulator}\label{subsec:full_algorithm}

\Cref{algo:simulator} shows the integration of the previous sections into the full PINNSim algorithm that is used for the computation of a time step.   
\begin{algorithm}
\caption{PINNSim - single time step}

{\small \textbf{Require:}
$\bm{x}_0$, $t_0$, $t_{\max}$, $\Delta t$, $s$, $r$, $\Delta \Vparams^{\max}$, $k^{\max}$\\
\textbf{Initialise:} $k = 0, \Vparamsiter{0}, \Delta \Vparamsiter{0}$
\begin{algorithmic}[1]\label{algo:simulator}
\While{$\Delta \Vparamsiter{k} > \Delta \Vparams^{\max}$ and $k < k^{\max}$}
\For{component $i = 1,\dots, m$ \& query points $j = 1, .., s$}
\State Predict state with PINN $\xstatehati{i}(t_j, \xinitiali{i}, \Vparamsiteri{k}{i})$
\State Calculate component injections $\IcomplexComponentHati{i} = h_i(t_j, \xstatehati{i}, \Vparamsiteri{k}{i})$
\State Calculate contribution of $\IcomplexComponentHati{i}$ to \Jacobian{}
\EndFor {\bf end}
\For{query points $j = 1, \dots, s$}
\State Calculate network injections $\IcomplexNetworkHat(t_j, \Vparamsiter{k})$
\State Calculate contribution of $\IcomplexNetworkHat$ to \Jacobian{} 
\EndFor {\bf end}
\State Assemble \Jacobian{} and solve $\left(\bm{J}^\top \bm{J}\right) \Delta \Vparamsiter{k+1} = - \bm{J} \residualError$
\State Update iteration $\Vparamsiter{k+1}=\Vparamsiter{k} + \Delta \Vparamsiter{k}$, $k = k + 1$
\EndWhile {\bf end}
\State \Return Trajectory across time step $t, \xstatehat{}(t), \Vcomplexhat{}(t)$ with \Vparamsiter{\text{final}}

\end{algorithmic}}
\end{algorithm}
Its accuracy \revision{is dependent}{depends} on the approximation quality of $\xstatehat{}(t)$ and $\Vcomplexhat(t)$ and on the tolerance settings for $\Delta \Vparams$ and the maximum number of iterations $k^{\max}$. Therefore, requirements on the resulting tolerance limit the suitable time step size $\Delta t$. By repeatedly applying \cref{algo:simulator}, we obtain a time-stepping scheme that then allows the simulation of dynamics beyond $\Delta t$.

\section{\revision{Case study}{Numerical experiments}}\label{sec:case_study}

\revision{This section describes the modelling of the power system dynamics and the implementation of PINNSim.}{This section presents the setup of the numerical experiments to illustrate a proof of concept for PINNSim. The chosen dynamical model describes power system dynamics, we note though, that we use a simplified model as the focus lies on the numerical aspects rather than obtaining insights into the transient phenomena.}

\subsection{Power system modelling}
As an example for a dynamic component, we consider a two-axis generator model as modelled in \cite{sauer_power_1998}.
\begin{subequations}\label{eq:generator_model}
\begin{align}
     \begin{bmatrix} \scriptstyle T'_{do} \\ \scriptstyle T'_{qo} \\ \scriptstyle 1 \\ \scriptstyle 2H \end{bmatrix} \frac{d}{dt}& \begin{bmatrix} \scriptstyle E'_q \\ \scriptstyle E'_d\\ \scriptstyle \delta \\ \scriptstyle \Delta \omega \end{bmatrix} =\begin{bmatrix} \scriptstyle -E'_q - (X_d - X'_d) I_d + E_{fd} \\ \scriptstyle -E'_d + (X_q - X'_q) I_q \\ \scriptstyle \omega_s \Delta \omega \\ \scriptstyle P_m - E'_d I_d - E'_q I_q - (X'_q - X'_d) I_d I_q - D \Delta \omega \end{bmatrix}\label{eq:generator_model_f}\\
    \begin{bmatrix}
        I_d \\ I_q
    \end{bmatrix} &= \begin{bmatrix}
        R_s & -X'_q \\
        X'_d & R_s
    \end{bmatrix}^{-1} \begin{bmatrix}
        E'_d - V \sin{(\delta - \theta)}\\
        E'_q - V \cos{(\delta - \theta)}
    \end{bmatrix}\label{eq:generator_model_h1}\\
    \Bar{i}^C &= (I_D + j I_Q) = (I_d + j I_q) e^{j(\delta - \pi/2)}\label{eq:generator_model_h2}.
\end{align}
\end{subequations}
\labelcref{eq:generator_model_f} corresponds to \labelcref{eq:f_update_power_system} and \labelcref{eq:generator_model_h1,eq:generator_model_h2} to \labelcref{eq:h_update_power_system}. For this study, we simplify the model above to a classical machine model by setting the reactances to $X'_q = X'_d$ and $X_q = X'_d$ and then finding the integral manifold such that the internal voltages $E'_q$ and $E'_d$ remain constant at $E'_{q0}$ and $E'_{d0}=0$. For more details we refer to \cite{sauer_power_1998}. Now, the rotor angle $\delta$ and the frequency deviation $\Delta \omega$ form the state \xstatei{i}, the magnitude $V$ and angle $\theta$ of the terminal voltage form \Vcomplexi{i}, and the mechanical power $P_m$ and the excitation voltage $E_{fd}$ form the control input \ustatei{i}. More detailed \revision{}{components} models could include higher order electro-mechanical modes, governor dynamics for $P_m$ and exciter dynamics $E_{fd}$. Similarly, inverter-based resources or voltage dependent loads could be included.

\revision{To illustrate PINNSim, we consider}{The results in \cref{sec:results} will demonstrate how PINNSim can effectively increase the allowable time step size. We observe this characteristic already for the following simple setup. We consider} the IEEE 9-bus system described in \cite[pp.~164--167]{sauer_power_1998} with three generators (all modelled as classical machines as above) with parameters from \cref{tbl:generator_parameters}. The initial conditions \xinitiali{i} and control inputs \ustatei{} are determined from assuming an equilibrium state for the load flow case in \cite{sauer_power_1998}. To perturb the system, we reduce the mechanical power $P_m$ of generator 1 to 50\% of its initial value and then observe the resulting trajectory\revision{}{\footnote{Here, we apply a disturbance that leaves the power flow at the time of the event unchanged. In contrast, short-circuits and topology changes alter the power flow as the algebraic variables change instantaneously. The resulting jumps can be challenging for simulators. Treating these cases with PINNSim will be addressed in future work.}}.
\begin{table}[!th]
\renewcommand{\arraystretch}{1.2}
\caption{Generator parameters and set points (all in \si{\pu})}
\label{tbl:generator_parameters}
\centering
\begin{tabular}{cccccccc}
\toprule
Gen. & $H$ & $D$ & $X_d$ & $X'_d$ & $R_s$ & $P_m$ & $E_{fd}$\\\midrule
1 & 23.64 & 2.364 & 0.146 & 0.0608 & 0.0 & 0.71 & 1.08\\
2 & 6.4 & 1.28 & 0.8958 & 0.1969 & 0.0 & 1.612 & 1.32\\
3 & 3.01 & 0.903 & 1.3125 & 0.1813 & 0.0 & 0.859 & 1.04\\ \bottomrule
\end{tabular}%
\end{table}

\subsection{Implementation and NN training}
\revision{
The entire simulator is implemented in PyTorch \cite{paszke_pytorch_2019} as we utilise \gls{AD} for the computation of the Jacobian \Jacobian{} and it enables the training of the \gls{PINN} models. For each generator we used 2500 simulated points in the training dataset $\mathcal{D}$ and 5000 collocation points. Each \gls{PINN} consists of 2 hidden layers with 32 neurons, uses a $\tanh{}$ activation function, and is trained for a range of possible voltage parametrisations \Vparamsi{i} with $r=2$, initial conditions \xinitiali{i}, and prediction time steps $\Delta t \in [0, 0.3] \si{\second}$, which results in a nine-dimensional input space. We train, using a L-BFGS optimiser, for 2000 epochs. The published code base \cite{stiasny_publicly_2022} provides a detailed overview of the dataset creation setup and the training procedure.
The numerical simulations of the system of \glspl{DAE} that serve as ground truth are implemented using Assimulo \cite{andersson_assimulo_2015}. As a comparison to PINNSim, we implement a trapezoidal integration scheme, see \cite{milano_power_2010}.
}{
The entire simulator is implemented in PyTorch \cite{paszke_pytorch_2019} as we require the functionality of \gls{AD} for the computation of the Jacobian \Jacobian{} and the training of the \gls{PINN} models. Each \gls{PINN} consists of three hidden layers with 32 neurons and applies a $\tanh{}$ activation function. We employ a Xavier-normalised initialisation \cite{glorot_understanding_2010} for the \gls{PINN} parameters and subsequently optimise them using a L-BFGS optimiser \cite{liu_limited_1989} for 2000 epochs.
For each generator the training dataset $\mathcal{D}$ comprises 2500 simulated points, 500 of which form a validation dataset. The collocation dataset $\mathcal{D}_c$ has size 5000. For all datasets we sample from the input domain which consists of the prediction time step $\Delta t$, the initial condition \xinitiali{i}, and a voltage parametrisations \Vparamsi{i} with $r=2$. The bounds of the resulting nine-dimensional input domain are shown in \cref{tbl:dataset_parameters}. They are chosen such that all conditions encountered in the simulation are covered. We use the equilibrium values $\delta_i^{eq}, \theta_i^{eq}, V_i^{eq}$ from the power flow solution in the definition of the domain, but it can also be defined without this information.}

\revision{}{
The numerical simulations of the system of \glspl{DAE} that serve as ground truth are performed using Assimulo \cite{andersson_assimulo_2015}. As a direct comparison to PINNSim, we implement a trapezoidal integration scheme, see \cite{milano_power_2010}. The timing of the simulations is conducted on a AMD Ryzen 7 PRO (1.9 GHz, 8 cores) and 16GB RAM with the implementation of PINNSim that is publicly available \cite{stiasny_publicly_2022}.
}
\begin{table}[!h]
\renewcommand{\arraystretch}{1.2}
\caption{\revision{}{Input domain of the training dataset}}
\label{tbl:dataset_parameters}
\centering
\begin{tabular}{ccc}
\toprule
$\Delta t$ & $\delta_{0,i} - \delta_i^{eq} $ & $\Delta \omega_{0,i}$ \\\midrule
$[0.0, 0.3]\,\si{\second}$ &  $[-\pi, \pi]\,\si{\radian}$ & $[-0.86, 0.86]\,\si{\hertz}$\\ \bottomrule
\vspace{-0.2cm}\\
\toprule
$\theta_{0,i} - \theta_i^{eq}$& $\theta_{1,i}$ & $\theta_{2,i}$ \\\midrule
$[-\pi, \pi]\,\si{\radian}$ & $[-0.3, 0.3]\,\si{\radian\per\second}$ & $[-0.8, 0.8]\,\si{\radian\per{\second^2}}$ \\ \bottomrule
\vspace{-0.2cm}\\
\toprule
$V_{0,i}- V_i^{eq}$& $V_{1,i}$ & $V_{2,i}$ \\
\midrule
$[- 0.1, 0.3]\,\si{\pu}$ & $[-0.4, 0.4]\,\si{\pu\per\second}$ & $[-0.5, 0.5]\,\si{\pu\per{\second^2}}$ \\ \bottomrule
\end{tabular}%
\end{table}

\section{Results}\label{sec:results}

The potential for accelerating \glspl{TDS} with PINNSim relies on increasing the time step size and thereby reducing the total number of time steps compared to established methods such as the trapezoidal method. We demonstrate this behaviour in \cref{subsec:full_simulation} on the test case. In \cref{subsec:single_step}, we then illustrate what allows PINNSim these larger time steps and describe in \cref{subsec:computational_cost} how the involved computational cost compare.

\subsection{Accuracy of PINNSim for a full TDS}\label{subsec:full_simulation}
\begin{figure}[!t]
    \centering
    \includegraphics[width=\linewidth]{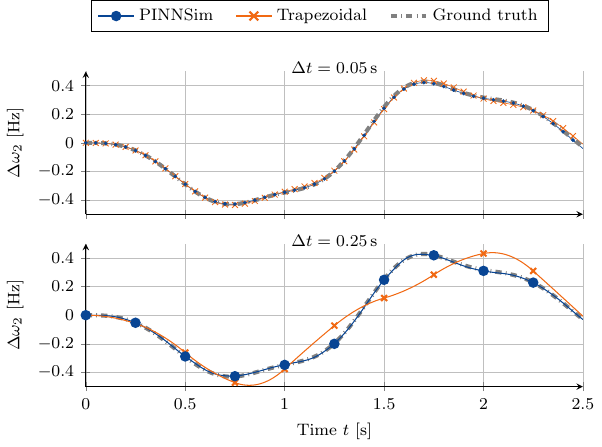}
    \caption{\revision{}{The panels show the simulated trajectories of the frequency deviation at machine 2 $\Delta \omega_2$ with a trapezoidal and PINNSim time stepping scheme for two time step sizes. The markers indicate the values at the time steps. The curves within a time step stem from the prediction of the PINNs for PINNSim and from a quadratic interpolation for the trapezoidal integration. The ground truth stems from a highly accurate integration scheme.}}
    \label{fig:trajectory_analysis}
\end{figure}

First, we consider a simulation of the described test case over \SI{2.5}{\second}. \revision{\Cref{fig:trajectory_analysis} shows in the upper panels the resulting trajectory of the frequency deviation at machine 2, i.e., $\Delta \omega_2$.}{PINNSim will return trajectories for all states $\xstatei{i}(t)$ and voltages $\Vcomplexi{i}(t)$. For clarity of the presentation, we will subsequently only focus on one state, namely the frequency deviation at machine 2, i.e., $\Delta \omega_2$. \Cref{fig:trajectory_analysis} shows the resulting trajectory of $\Delta \omega_2$. The grey dash-dotted line represents the ground truth solution of the power system dynamics which stems from the Assimulo solver with a tolerance setting of $10^{-12}$. The time-stepping schemes, i.e., PINNSim and trapezoidal method, return the values at the end of each time step, indicated by the markers. To obtain the intermediate values we either query the PINNs (for PINNSim) or apply a quadratic interpolation (for the trapezoidal method). While PINNSim and the trapezoidal rule accurately capture the dynamics for a time step size of $\Delta t = \SI{0.05}{\second}$, the trapezoidal rule fails to track the evolution for a larger time step size of $\Delta t = \SI{0.25}{\second}$. In contrast, PINNSim captures the state evolution accurately.}

\begin{figure}[!b]
    \centering
    \includegraphics[width=\linewidth]{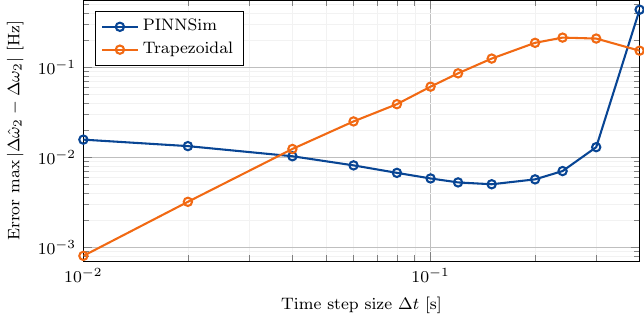}
    \caption{\revision{}{Comparison of the maximum error of $\Delta \omega_2$ over a trajectory of $\SI{2.4}{\second}$ for a range of time step sizes $\Delta t$.}}
    \label{fig:trajectories_error}
\end{figure}
We test this behaviour for more time step sizes and \revision{plot the maximum error over this time interval in \cref{fig:trajectories_error}}{plot in \cref{fig:trajectories_error} the maximum error within the simulation time interval}. The trapezoidal rule performs well for very small time step sizes, but the errors quickly become very large - for reference, the overall variation of $\Delta \omega_2$ is around \SI{0.8}{\hertz}. PINNSim leads to an almost constant error \revision{characteristics}{characteristic}, mostly with less than \SI{0.01}{\hertz} error. The slightly increasing error towards smaller time step sizes arises due to accumulating errors. The larger errors for $\Delta t > \SI{0.3}{\second}$ are expected, as we trained the \glspl{PINN} only up \revision{}{to} this time step size.

\subsection{Accuracy of PINNSim on a single time step}\label{subsec:single_step}

In the following, we consider a single time step and show the influencing factors on the accuracy. To obtain a range of system states, we simulate the test case for \SI{10}{\second} with an accurate solver and then extract 200 instances, i.e., every \SI{0.05}{\second}.

\subsubsection{Voltage parametrisation} 

We use these 200 instances as initial values to test the performance of PINNSim on a single time step for different values of $\Delta t$, the voltage profile order $r$ and the number of query points $s$. 
\begin{figure}[th]
    \centering
    \includegraphics{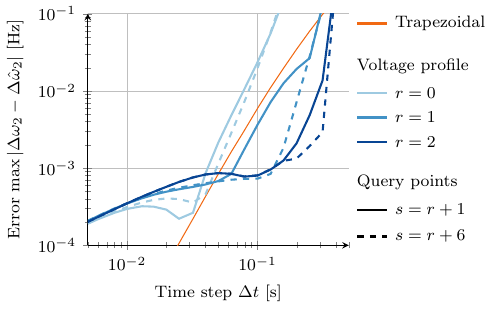}
    \caption{Maximum error for single time step predictions of different length $\Delta t$ starting from 200 different initial conditions. Comparison between the trapezoidal rule and PINNSim with different voltage scheme orders ($r$) and number of query points $s$. Larger $r$ and $s$, both improve the accuracy of PINNSim.}
    \label{fig:time_step_analysis}
\end{figure}
We evaluate these time steps with respect to the maximum error of $\Delta \omega_2$ at the end of the time step. \Cref{fig:time_step_analysis} reports the results. A main observation is that PINNSim with $r=1$ (linear voltage evolution) and $s=r+1$ has a comparable relationship between accuracy and time step size as the trapezoidal rule. However, when using PINNSim with $r=2$, the time step size can be significantly increased without incurring much larger errors. For very small time step sizes PINNSim offers no benefits over the trapezoidal rule as the error decreases slower when reducing $\Delta t$. We can improve this characteristic for PINNSim by evaluating \cref{eq:objective_approximation} on more points, i.e., \revision{choose}{choosing} a larger value for $s$, shown as the dashed lines. The improved performance, in particular for $r=1$, originates from the better approximation of \cref{eq:objective_voltage_update_exact}. However, for large time steps, here for around $\Delta t > \SI{0.2}{\second}$, we additionally require $r=2$ for accurate predictions; the voltage evolution becomes too non-linear, hence, the linear voltage approximation ($r=1$) is insufficient. For $r=2$, the limitation in time step size arises as \revision{}{the} training domain of the \glspl{PINN} was restricted to $\Delta t \leq \SI{0.3}{\second}$.

\subsubsection{PINN accuracy} 
\begin{figure}[ht]
    \centering
    \includegraphics[width=0.95\linewidth]{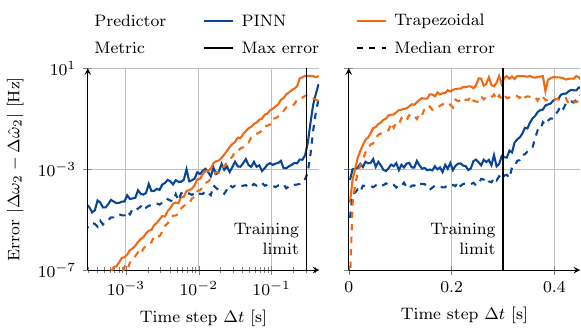}
    \caption{Error characteristics of the PINN for machine 2 on a test dataset. Both plots show the same results but on a logarithmic x-axis (left) and a linear x-axis (right) to highlight the accuracy of PINNs over large time steps.}
    \label{fig:error_characteristic_PINN}
\end{figure}

The performance of PINNSim relies in large parts on how well each \gls{PINN} approximates the integration in \cref{eq:x_integral} for a given voltage profile $\Vcomplexihat{i}(t, \Vparamsi{i})$. \Cref{fig:error_characteristic_PINN} illustrates the results on a test dataset of 4000 points for generator 2. Both panels show the same results, i.e., the maximum and median error on the test dataset, but on a logarithmic and linear x-axis. The logarithmic axis clearly shows that for small time step sizes, the error of \glspl{PINN} decreases as we included a time dependency in the final layer \cref{eq:NN_output}. The linear x-axis gives a more intuitive understanding of how much further \glspl{PINN} can predict the dynamics accurately. Only when reaching the limit of the training domain, the accuracy deteriorates. Revisiting the error plots in \revision{\cref{fig:trajectory_analysis,fig:time_step_analysis}}{\cref{fig:trajectories_error,fig:time_step_analysis}}, we can find that it is the particular error characteristic of \glspl{PINN} that allows the large time steps of PINNSim and why it outperforms the trapezoidal method. 

\subsection{Computation cost of a time step}\label{subsec:computational_cost}
The larger time step sizes of PINNSim come at \revision{increasd}{an increased} computational cost compared to the trapezoidal rule. \revision{Providing a fair comparison based on optimised implementations will be future work, instead, we will here focus on the main cost of PINNSim from a perspective of scalability. 
}{We first analyse this cost of PINNSim from a perspective of scalability and then provide a comparison of the total run-time based on the implementation of this proof of concept.} 

In \cref{algo:simulator}, the evaluation of the inner loop (lines 2-8) requires the calculation of network and component current injections \IcomplexNetworkHat{}, \IcomplexComponentHat{} and their sensitivity to \Vparams{}. For the network currents, this requires merely matrix-vector products and for the component currents the \revision{evaluations}{evaluation} of the \glspl{PINN}. A single evaluation (pass) of a \gls{PINN} requires on the order of \SI{1}{\micro\second}, e.g., to compute the current injection\revision{, and the}{. The} sensitivity calculations \revision{requires}{require} only two additional passes thanks to \gls{AD}. As all \glspl{PINN} could be evaluated in parallel, these computations can easily be scaled to large system. The \say{expensive} computation arises in the solution of the linear system of equations in line 9 of \cref{algo:simulator}, but its sparsity and known structure can be exploited. A closely related calculation is required for the trapezoidal rule.

\begin{figure}[!t]
    \centering
    \includegraphics[width=0.95\linewidth]{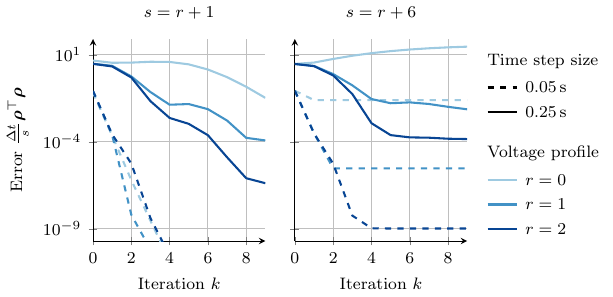}
    \caption{Convergence of the objective value in \cref{eq:objective_approximation} for different number of query points $s$, voltage profile orders $r$ and time step sizes $\Delta t$.}
    \label{fig:convergence_analysis}
\end{figure}

The overall cost then scales linearly with the number of iterations $k$ per time step, i.e., line 1-10 in \cref{algo:simulator}, until convergence is reached. We show in \cref{fig:convergence_analysis} the value of the objective \cref{eq:objective_approximation} over the iterations for two time step sizes and different values of $r$. In the left panel, the least square problem is fully determined ($s=r+1$), hence the objective value continues to decrease. In the right panel, the five additional query points render the problem over-determined, hence the algorithm converges at a non-zero objective value. In either case, we observe that larger time step sizes require more iterations, while the order of the voltage profiles primarily affects the magnitude of the objective value and less its convergence.

\revision{}{
Lastly, we provide a comparison of the absolute run-times for the trapezoidal method and PINNSim in \cref{tbl:simulator_results}. We report the run-time for different combinations of the time step size $\Delta t$, the number of iterations $k$ per time step, and the number of query points $s$ for PINNSim. We observe that one faces a trade-off between speed and accuracy with either method. The best values are highlighted and we note that the time step size and the number of iterations $k$ have the biggest influence on this trade-off. We want to stress though, that neither method is optimised for speed, hence the presented results are only indicative for the implementation in \cite{stiasny_publicly_2022}.
\begin{table}[!ht]
\renewcommand{\arraystretch}{1.2}
\caption{\revision{}{Comparison of total run-time for simulating $\SI{2.4}{\second}$}}
\label{tbl:simulator_results}
\centering
\begin{tabular}{cccccc}
\toprule
Simulator & $\Delta t \,[\si{\second}]$ &  $k$ & $s$ & run-time $[\si{\second}]$ & $\max | \omega_2 - \hat{\omega}_2| \, [\si{\hertz}]$ \\ \midrule
\multirow{3}{*}{Trapezoidal} & \multirow{2}{*}{$0.04$} & $1$ & $-$ & $\bm{0.47}$ & $1.34 \times 10^{-2}$\\
& & $2$ & $-$ &  $0.90$ & $\bm{1.24 \times 10^{-2}}$\\
& $0.06$ & $2$ & $-$ & $0.59$ & $2.50 \times 10^{-2}$\\ \midrule
\multirow{5}{*}{PINNSim} & \multirow{3}{*}{$0.20$} & $5$ & $3$ & $0.72$ & $1.59 \times 10^{-2}$\\
& & $5$ & $7$ & $0.78$ & $0.98 \times 10^{-2}$\\
& & $6$ & $3$ & $0.88$ & $\bm{0.88 \times 10^{-2}}$\\
& \multirow{2}{*}{$0.24$} & $5$ & $4$ & $\bm{0.62}$ & $2.52 \times 10^{-2}$\\
& & $5$ & $7$ & $0.67$ & $1.48 \times 10^{-2}$\\ 
\bottomrule
\end{tabular}%
\end{table}
}

As stated in \cite{stiasny_physics-informed_2023,stiasny_physics-informed_2023-1}, the cost of training the \glspl{PINN} has to be considered as well. Improving the efficiency and performance of the training, \revision{}{e.g., through advanced \gls{PINN} architectures or hyperparameter tuning,} however, can be decoupled from the analysis of PINNSim. 

\section{Discussion}\label{sec:discussion}

The presented results shall serve as a proof of concept of PINNSim as a novel time stepping simulator. We demonstrated that PINNSim outperforms the trapezoidal rule on the metric of the allowable time step size. The following describes four aspects that are necessary to turn PINNSim from a proof of concept into a fully fledged simulator for power system dynamics and to realise the potential acceleration of \glspl{TDS}.

\begin{enumerate}
    \item \textit{Speed}: Numerical methods require a high level of optimisation in the implementation to become competitive. For PINNSim this concerns primarily the optimisation of the calculation of the residual \residualError{}, the Jacobian \Jacobian{}, and the solution of the sparse linear system in \cref{eq:residual_least_square}. The focus should therefore lie on controlling memory allocations, utilising parallelisation, and exploiting sparsity patterns for these computations.
    \item \textit{Accuracy}: The accuracy of PINNSim hinges around accurately learned \glspl{PINN}. We envision that this process can be highly standardised, so that \glspl{PINN} can be trained reliably to high accuracy and with desirable error characteristics for a wide range of dynamic components.
    \item \textit{Scale}: The analysis of the computational complexity of PINNSim suggests its scalability but it remains to be shown in practice. Furthermore, we need to investigate the accuracy and convergence properties of the voltage update scheme for larger systems.
    \item \textit{\revision{Applications}{Power system scenarios}}: In this work, we have considered the dynamics of classical machine models under a set-point change. Future work should include more detailed dynamical models and apply PINNSim to short-circuits, topology changes, and other numerically more demanding setups \revision{.}{to demonstrate its adequacy for the simulation of power system dynamics.}
\end{enumerate}

\section{Conclusion}\label{sec:conclusion}

With PINNSim, we introduced a novel approach for time-domain simulations that allows the integration of \glspl{PINN}. As a result we achieve accurate results with significantly larger time steps than with the common trapezoidal method. Hence, we require fewer time steps which could result in a significant acceleration of time-domain simulations in power systems. By design, we only require the training of \glspl{PINN} for single dynamical components and thereby enable the scalability of PINNSim to large systems. Additionally, all calculations at run-time are highly scalable and parallelisable, necessary requirements to develop PINNSim into a powerful simulator.

\bibliographystyle{IEEEtran}
\bibliography{bstcontrol.bib,references_offline.bib}

\end{document}